\documentstyle[a4,12pt]{article}
\def\be{\begin{equation}}
\def\ee{\end{equation}}
\def\bq{\begin{eqnarray}}
\def\eq{\end{eqnarray}}
\begin{document}

\thispagestyle{empty}
\setcounter{page}{0}
\begin{flushright}
\bf{LMU-01/94}\\
March 1994
\end{flushright}
\vspace*{\fill}
\begin{center}
{\Large\bf
Calculation of Chirality Violating \\
Proton Structure Function h$_1$(x) in QCD}\\
\vspace{2em}

\large
{\bf B.L.Ioffe}\\
\vspace{1em}
{\small Institute of Theoretical and Experimental Physics \\
117259 Moscow, Russia }\\
\vspace{2em}

\large
{\bf A. Khodjamirian$^*$ }\\
\vspace{1em}
{\small Sektion Physik der Universit\"at M\"unchen\\
D-80333 M\"unchen , Germany\\
and\\
Yerevan Physics Institute, 375036 Yerevan, Armenia } \\

\end{center}
\vspace*{\fill}

\begin{abstract}
The twist-two chirality violating proton structure function $h_1(x)$
measurable in the polarized Drell-Yan process is calculated by means
of
QCD sum rules at intermediate $x$, $0.3 < x < 0.7$ and $Q^2 \approx
5-10
GeV^2$.

\end{abstract}

\vspace*{\fill}

\begin{flushleft}
\noindent$^*$ {\it Alexander von Humboldt Fellow}\\
\end{flushleft}

\newpage
\section{Introduction}

\vspace{5mm}
As is well known, all structure functions of the lowest twist-two,
$F_1(x), F_2(x), g_1(x)$ which are measured in the deep-inelastic
lepton-nucleon scattering, conserve chirality. Ralston and Soper
\cite{1}
first demonstrated that besides these structure functions, there
exists the
twist-two chirality violating nucleon structure function $h_1(x)$.
This
structure function does not manifest itself in the deep inelastic
lepton-hadron scattering, but can be measured in the Drell-Yan
process with
both beam and target transversely polarized. The reason of this
circumstance is the following. The cross section of the deep
inelastic
electron(muon)-hadron scattering is proportional to the imaginary
part of
the forward virtual photon-hadron scattering amplitude. At high
photon
virtuality the quark Compton amplitude dominates, where the photon is
absorbed and emitted by the same quark (Fig.1a) and the conservation
of
chirality is evident. The cross section of the Drell-Yan process can
be
represented as an imaginary part of the Fig. 1b diagram. Here virtual
photons interact with different quarks and it is possible, as is shown
in
Fig.1b, that their chiralities are also different. It is clear from
Fig.1b,
that chirality violating amplitude in Drell-Yan processes is not
suppressed
at high $Q^2$ in comparison with chirality conserving ones and,
consequently, corresponds to twist two.
However, this amplitude
corresponding to target spin flip, has no parton interpretation in the
standard helicity basis. Nevertheless, $h_1(x)$ can be interpreted as a
difference
of quark densities with eigenvalues $+1/2$ and $-1/2$ of the transverse
Pauli-Lubanski operator $\hat{S}_{\perp}\gamma_5$
in the transversely polarized
proton as it is explained in details by Jaffe and Ji in
\cite{2}.

Until now, there is no experimental data on the chirality violating
nucleon
structure function $h_1(x)$. Besides \cite{1,2} the theoretical study
of
this structure function ( under the name $\Delta_1q(x)$ ) have been
performed also by Artru and Mekhfi \cite{AM}.
The role of $h_1(x)$ in
factorization of a general hard process with polarized particles was
investigated by Collins \cite{Collins}. The first attempt to
calculate
$h_1(x)$ was carried out by Jaffe and Ji \cite{2} by means of the
bag model.

In this paper we calculate the proton structure function $h^p_1(x)$
by
means of the QCD sum rule approach. The idea of the method is the
following
\cite{3,4}. Consider the four-point vacuum correlator corresponding
to the
amplitude of forward scattering of the current $\eta(x)$ with proton
quantum
numbers on external currents $j_1(x), j_2(x)$.
(In the  case of the structure function $F_2(x)$ considered earlier
\cite{3,4} ~~~ $j_1(x), j_2(x)$ were electromagnetic or
weak currents). Let the momenta, corresponding to the currents
$\eta(x)$ and
$j(x)$ be $p$ and $q$. As was shown in \cite{3,4}, the imaginary part
of the
forward scattering amplitude is determined by small distances
in $t$-channel
if $p^2$ and $q^2$ are negative and large enough, $\mid p^2 \mid,
\mid q^2
\mid \gg R^{-2}$, (R is the confinement radius) and the Bjorken
scaling
variable $x = Q^2/2 \nu, ~~~ Q^2 = -q^2, ~~ \nu = p \cdot q$ is not
close to the
boundary values $x=0$ and $x=1$.  Therefore, to calculate the
amplitude of
interest we may use the operator product expansion (OPE) method
accounting
the vacuum expectation values (v.e.v.) of various operators. If, in
addition, we suppose $\mid p^2 \mid \ll \mid q^2 \mid$ and restrict
ourselves to the first term in expansion over $p^2/q^2$, only
twist-two
structures will be retained. In order to obtain the desired
proton structure
function we use the standard procedure of the QCD sum rule approach.
The amplitude calculated in QCD
is equated to the contribution of physical states, the proton and
the excited states with proton quantum numbers, by means of the
dispersion relation in $p^2$. The proton contribution we
are interested in, is then separated by applying the Borel
transformation
in $p^2$ which suppresses the contribution of excited  states.

It is important to check that the results satisfy necessary
conditions of
QCD sum rule calculations: 1) the OPE is converging i.e., the
the highest accounted term in OPE is well less than the
first term$(s)$ of OPE and 2) in the phenomenological part of the sum
rule the contribution of excited states does not surpass the one of
the proton.
These requirements determine the domain of $x$, where the results
are valid. The value of $Q^2$ must be much larger than the proton mass $m$
squared,$ ~~ Q^2 \gg m^2$,
but cannot be very large, since the leading logarithmic corrections
in QCD are not accounted in our approach. So, we expect that our results are
valid at $Q^2 \sim 5-10 GeV^2$.

It should be emphasized that calculation of various proton structure functions
by means of QCD sum rules contains no experimental inputs. Moreover, there
are no
new parameters at all. The values of vacuum condensates determining the OPE
are fixed from QCD sum rules for various two-point
correlators, whereas the parameters of the proton channel, namely the
transition constant of proton into the quark current
and the value of the continuum threshold are taken
from the analysis of the sum rules for the proton mass \cite{5}.
The price for model independence is a limited domain of
$x$ where the calculated structure function is reliable.
Thus, for the valence $u(x)$ distribution
calculated in \cite{4} the region of validity
$0.2 < x <0.7$ was established.
Calculated by the same method, the $u$ quark contribution to the
proton spin structure function $g_1^p(x)$ \cite{g1}
has an even narrower interval of validity $0.5 < x < 0.7$. The criteria
determining these intervals as well as the accuracy of
the calculation are well defined from comparing nonleading and
leading terms of OPE, as explained above.
A comparison with experiment shows that one can rely on this
method of calculation within the domain of validity.
A very nontrivial evidence of that is the accordance of the
calculated value of $g_1^p(x)$  with recent experimental data
( see detailed discussion below).

In carrying out this program for calculation of $h_1(x)$ some
specifics
arise. In refs.\cite{1,2} the structure function $h_1(x)$ was
represented in
terms of the matrix elements of the operators on the light-cone.
In our
calculation we start from the forward scattering amplitude on the
proton. In
Sec. 2 the forward scattering amplitude corresponding to the light
cone
definition of $h_1(x)$ is constructed. The four-point current
correlator
calculated in QCD contains many spin-tensor structures and it is
necessary
to find such structure, which, if saturated by the proton state,
gives the
desired function $h_1(x)$. This is done in Sec.3.

In sec.4 we calculate the OPE of the four-point current correlator.
Here
the most important difference between the function $h_1(x)$ and other
twist-two structure functions appears: since $h_1(x)$ violates
chirality,
the OPE in the chosen spin-tensor structure starts from the
operator of
dimension 3 - the quark condensate, unlike the case of other
twist-two structure
functions, where it starts from the zero dimension unit operator. In
addition to the main term of OPE we calculate the next term
proportional to
the mixed quark-gluon condensate of dimension 5.

In Sec.5 we obtain the sum rule, perform numerical
analysis and give the
results of the $h_1(x)$ determination. Our conclusions are
presented in
Sec.6.

\vspace{5mm}
\section{The connection of light-cone and forward
scattering amplitude definitions of the
structure function $h_1(x)$}.

The structure function $h_1(x)$ was defined in ref.\cite{2} by
Fourier
transformation of the operator product on the light cone between
proton
states

$$ i \int\limits \frac{d \lambda}{2 \pi} e^{i \lambda x}
< p,s\mid \bar{\psi}(0)
\sigma_{\mu \nu} \gamma_5 \psi (\lambda n) \mid p,s > =$$
$$2 [h_1(x,q^2)(s_{\bot \mu} p_\nu - s_{\bot \nu}
p_\mu)+
h_L(x,q^2) m^2(p_\mu n_\nu -$$
\begin{equation}
p_\nu n_\mu)(s \cdot n) +
h_3(x,q^2)m^2 (s_{\bot \mu} n_\nu - s_{\bot \nu} n_\mu)]
\end{equation}
Here $n$ is a light cone vector of dimension of $(mass)^{-1}$,
$n^2 = 0$,~$n^+ = 0$,
$ p\cdot n = 1$,~~$p$ and $s$ are the proton momentum
and spin vectors:
$p^2 = m^2$, $s^2 =-1$,~~$p \cdot s = 0$,~ and
$ s=(s \cdot n)p +(s \cdot p) n +s_{\bot} $.
We choose in this section the
proton and virtual photon c.m.system and put
$q = \{ q_0,0,0,E \}$, $p =
\{E,0,0,-E \}$, ~~ $n = \{1/E,0,0,1/E \}/2$, where
$E$ is the virtual photon
momentum in this system and the terms of order
$m^2/E^2$ are neglected.

Let us demonstrate now that $h_1(x)$ can also be defined as a
matrix element
between proton states for forward scattering of the axial
current with its
transition into scalar current $p A \rightarrow pS$ plus
crossing matrix
element \footnote {We are indebted to R.Jaffe for the idea of this
definition of $h_1(x)$.}

\begin{equation}
T_{\mu}(p,q,s,s)= i \int\limits d^4x e^{iqx} < p,s \mid (1/2)
T \{ j_{\mu
5}(x),
j(0) + j(x),  j_{\mu 5}(0) \} \mid p,s>
\end{equation}
where
$j_{\mu 5}(x)$ and $j(x)$ are axial and scalar currents. Supposing
approximate conservation of axial current, $T_{\mu}(p,q,s,s)$ can be
represented as

\begin{equation}
T_{\mu}(p,q,s,s)=(s_{\mu} - \frac{q \cdot s}{q^2} q_{\mu})
\tilde{h}_1(x,q^2) +
\end{equation}
$$+(p_{\mu}-\frac{\nu q_{\mu}}{q^2})(q \cdot s) l_1(x,q^2) +
\epsilon_{\mu\nu
\lambda\sigma} p_{\nu} q_{\lambda} s_{\sigma}
(q \cdot s) l_2 (x,q^2)$$
(only spin-dependent terms are retained).
We will prove that

\begin{equation}
h_1(x,q^2) = - \frac{1}{\pi} Im \tilde{h}_1 (x,q^2)
\end{equation}
up to higher twist terms.

Until now we did not specify the flavors of axial and
scalar currents.
Strictly speaking, due to the anomaly,
the representation (3) is correct only
for flavor octet axial current. For the singlet axial current
additional terms appear in (3). But, as we shall see, in proton the
contribution of $u$-quarks to $h_1(x)$ is much larger than
of $d$-quarks.
Therefore, there is no matter whether we consider $\bar{u}u$
current, as we
shall do, or $\bar{u}u-\bar{d}d$ current, where anomaly does not
contribute
and (3) is correct up to the terms of order $m^2/Q^2$.
In any case, the
anomaly contribution is suppressed by $\alpha_s$, what is beyond the
accuracy of our results. For these reasons, we use eq.(1) for the
case of
one flavor.

As is well known, at high $Q^2$ the imaginary part of
the forward scattering
amplitude like (2) is dominated by the
light cone region in space-time.
Using the expression for quark propagator near light cone

\begin{equation}
< 0\mid T\{ \psi_{\alpha}(x),
\bar{\psi}_{\beta}(0) \} \mid 0> = -\frac{i}{4
\pi^2}(\gamma_{\mu})_{\alpha \beta} \frac{\delta}{\delta x_{\mu}}
\frac{1}{x^2-i \varepsilon}
\end{equation}
and substituting it into (2) we obtain

\begin{equation}
Im T_{\mu} = -\frac{1}{4 \pi} \int d^4 x e^{iqx} < p,s \mid
\bar{\psi}(x)(\hat{x}
\gamma_{\mu} - \gamma_{\mu} \hat{x}) \gamma_5 \psi(0)
\mid p,s> \delta' (x^2)
\end{equation}
Using the light cone variables

\begin{equation}
t - z = \tau ~~~~(1/2)(t + z) = \lambda
\end{equation}
we can rewrite (6) as

\begin{equation}
Im T_{\mu} = - \frac{i}{2} \int d \lambda \int d \tau \int dx^2
e^{iE(\tau-2 \xi \lambda)}
\theta(2 \tau \lambda - x^2) \delta'(x^2) \times
\end{equation}
$$\times <p,s \mid \bar{\psi}(x)
\sigma_{\mu \nu} x_{\nu} \gamma_5 \psi(0)
\mid p,s >$$
where the notation $\xi$ was temporarily
accepted for the Bjorken scaling
variable and we used the relation

\begin{equation}
qx \approx E(t-z) + \frac{q^2}{2 E} t \approx E \tau
- 2 \xi E \lambda
\end{equation}
Integrating over $x^2$ and $\tau$ in (8) and putting on the
light cone

\begin{equation}
x_{\nu} = 2E \lambda n_{\nu}
\end{equation}
we get

\begin{equation}
Im T_{\mu} = -\frac{i}{4}\int \limits_{-\infty}^{\infty}
d \lambda' e^{i \lambda' \xi} <p,s \mid \bar{\psi}(0)
\sigma_{\mu \nu}
n_{\nu} \gamma_5 \psi(n \lambda') \mid p,s>,
\end{equation}
where $\lambda' = 2 E \lambda$. Equation (11) can be
compared with (1),
multiplied by $n_{\nu}$. The comparison of
terms, proportional to $s_{\mu}$,
gives eq.(4). It should be mentioned that instead of axial and scalar
currents one can also use a combination of
vector and pseudoscalar currents
$j_{\mu}(x) j_5(0) - j_5(x) j_{\mu}(0)$.

\vspace{5mm}
\section{Selection of spin-tensor structures}
\vspace{3mm}

Consider the four-point correlator

\begin{equation}
\Pi_{\mu}(p,q) = -i \int \limits
d^4 xd^4 y d^4 z e^{iqx+ip(y-z)}\times
\end{equation}
$$\times <0 \mid T \{\eta(y), 1/2(j_{\mu5}(x), j(0) + j(x), j_{\mu
5}(0)), \bar{\eta}(z) \} \mid 0>$$
where $\eta$ is the three-quark current with the
proton quantum numbers
\cite{5}

\begin{equation}
\eta = \varepsilon^{abc}(u^a C \gamma_{\lambda} u^b) \gamma_5
\gamma_{\lambda} d^c
\end{equation}
$a, b, c = 1,2,3$ are colour indices. We are interested in the proton
contribution to $\Pi_{\mu}(p,q)$, which is given by

\begin{equation}
\Pi^{(p)}_{\mu}(p,q) = \lambda^2_N \frac{1}{(p^2-m^2)^2}
\sum_{r,r'} v^r(p)
T^{(p)}_{\mu}(p,q,r,r')\bar{v}^{r'}(p)
\end{equation}
where $v^r(p)$ is the proton spinor with momentum
and polarization $r$,~~
$\lambda_N$ is the transition constant of proton into quark current

$$<0 \mid \eta \mid p,r > = \lambda_N v^r(p)$$
and $T^{(p)}_{\mu}(p,q,r,r')$ is nondiagonal in the proton spin matrix
element (2). It is convenient to represent
$T^{(p)}_{\mu}(p,q,r,r')$ in the
form

\begin{equation}
T^{(p)}_{\mu}(p,q,r,r') = \bar{v}^r(p)
\tilde{T}^{(p)}_{\mu}(p,q) v^{r'}(p),
\end{equation}
Here $\tilde{T}^{(p)}_{\mu} (p,q)$ is the matrix
element between proton
states before multiplication on proton spinors,
i.e. where $p^2 = m^2$, but
$\hat{p} \not= m$.

Since the axial and scalar currents enter symmetrically (2) and (12),
$\Pi_{\mu}(p,q)$ as well as $\tilde{T}^{(p)}_{\mu}(p,q)$ satisfy the
condition

\begin{equation}
F(q,p) = C^{-1} F^T(-q, -p)C, ~~~ F
= \Pi_{\mu}, ~~ \tilde{T}^{(p)}_{\mu}
\end{equation}
where $C$ is the charge conjugation matrix.
Accounting for (16) the general
structure of $\tilde{T}^{(p)}_{\mu}(p,q)$ is

\begin{equation}
\tilde{T}^{(p)}_{\mu}(p,q) = [ K( \hat{q}
\gamma_{\mu}-\gamma_{\mu} \hat{q}) + E(\hat{p} \hat{q} -
\hat{q} \hat{p})p_{\mu} +
Cp_{\mu}\hat{q} + C' q_{\mu} \hat{p}
\end{equation}
$$+ D(\hat{p} \hat{q} - \hat{q} \hat{p})q_{\mu}
+ L(\hat{p} \gamma_{\mu} -
\gamma_{\mu} \hat{p}) + M \gamma_{\mu} + R \hat{q} q_{\mu} + S
\hat{p}p_{\mu} ] \gamma_5$$
where $K,E,C,C',D,L,M,R,S$ are functions of
$\nu = p \cdot q, q^2$ and
$p^2=m^2$. $\Pi_{\mu}(p,q)$ has the same structure, in this case
$p^2$ being not equal to $m^2$.

The matrix element of $\tilde{T}^{(p)}_{\mu}(p,q)$
over the polarized proton
state with spin $s_{\mu}$ is given by

\begin{equation}
(1/2) Tr \{\tilde{T}^{(p)}_{\mu}(p,q)(1 -
\gamma_5 \hat{s})(\hat{p}+m) \}
 =-4 s_{\mu} [\nu K+L m^2 +(1/2) Mm] +...
\end{equation}
where only the terms proportional to $s_{\mu}$
are written explicitly. In
accord with (3) these terms are proportional to $h_1(x)$.
Substitute (15)
into (14) and use the relation

$$\sum_{r}v^r_{\alpha}(p) \bar{v}^r_{\beta}(p) =
(\hat{p} + m)_{\alpha
\beta}$$

We have
$$\Pi^{(p)}_{\mu} = \frac{\lambda^2_N}{(p^2-m^2)^2}
(\hat{p}+m) \tilde
T^{(p)}_{\mu}(p,q)(\hat{p}+m)=
\frac{\lambda^2_N}{(p^2-m^2)^2} \times$$

$$ \{ -\frac{4m}{\nu} [(\nu K +
L m^2 +(1/2) M m) + m^2q^2 (D +(1/2)
R)] p_{\mu} \hat{q}$$

\begin{equation}
+ 4m^3(D +(1/2) R) \hat{q} q_{\mu} \} \gamma_5 + ...
\end{equation}
where the terms at the structures
$p_{\mu} \hat{q} \gamma_5$ and $q_{\mu}
\hat{q}\gamma_5$ in the decomposition
(17) are separated. When deriving (19)
we used the relation $\Pi_{\mu} q_{\mu}=0$
following from conservation of
axial current. As is seen from (18),
(19) the combination of the structure
functions $\nu K + L m^2 + M m/2$, which is proportional
to $h_1(x)$ can be
found  by studying $\Pi^{(p)}_{\mu}(p,q)$ as a
coefficient function at the
structure  $p_{\mu} \hat{q} \gamma_5$ added by the
coefficient function at
the structure $q_{\mu} \hat{q} \gamma_5$, multiplied
by $q^2/\nu = -2x$. In
fact, the latter, as can be easily seen, and will be
proved by direct
calculation in Sec.4 corresponds to twist 4 and
can be neglected.

Therefore, the recipe of the calculation of
$h_1(x)$ is the following.
Calculate  \newline
$Im \Pi_{\mu}(p,q)$ defined by eq.(12) in QCD at $p^2 < 0$ using
OPE in $1/p^2$ and retaining the leading terms in $1/q^2$,
corresponding to
twist 2 amplitude. Separate the coefficient function at the structure
$p_{\mu} \hat{p} \gamma_5$. On the other side,
using the dispersion relation
in $p^2$ represent the same function through the contribution of the
physical states - the proton and excited states,
the latter approximated by
continuum. Equate two representations and apply
the Borel transformation to
both sides of the sum rule in order to enhance
the proton contribution and
improve the convergence of OPE. This
(up to some details) gives the desired
sum rule for determination of the proton
structure function $h_1(x)$.

\vspace{5mm}
\section{QCD calculation of four-point correlator}

\vspace{5mm}
We calculate the imaginary part in $s$-channel of the four-point
correlator $\Pi_{\mu}(p,q)$ (12). Since $\Pi_{\mu}(p,q)$
violates chirality,
the OPE starts from the term proportional
to quark condensate $<0 \mid
\bar{q} q \mid 0 >$. The corresponding diagrams are given in Fig.2.

Consider first the case, when the axial and scalar
currents are $u$-quark
currents (Fig.2a). The calculation of the diagram Fig.2a gives

$$Im \Pi^u_{\mu}(p,q) = -\frac{1}{4 \pi \nu}
<0 \mid \bar{u} u \mid 0 >
[ (-4 + 9x + $$
\begin{equation}
+ 2x ln \frac{-p^2x}{2\nu})p_{\mu} \hat{q} + q_{\mu} \hat{q} ]
\gamma_5
\end{equation}
(only the structures $p_{\mu} \hat{q} \gamma_5$ and
$q_{\mu} \hat{q}$ are
retained). After the Borel transformation in
$p^2$ the contribution of the
structure $q_{\mu} \hat{q} \gamma_5$ vanishes and we get

\begin{equation}
B_{M^2} Im \Pi^u_{\mu}(p,q) =
\frac{1}{2 \pi \nu}< 0 \mid \bar{u}u \mid
0 > M^2 x p_{\mu} \hat{q} \gamma_5
\end{equation}
where $M^2$ is the Borel parameter.

In the case, when the scattering proceeds on $d$-quark,
it is clear that the
corresponding amplitude (Fig.2b)
in the lowest twist depends only on $\nu$
and $q^2$ and is $p^2$ independent.
(Direct calculation gives the value
$-(1/12 \pi) < 0 \mid \bar{d}d \mid 0 >/\nu x$
for the term proportional to
$p_{\mu} \hat{q} \gamma_5$). Thus, it vanishes
after Borel transformation.
The nonvanishing contribution in the case
of scattering comes from diagrams
with additional gluon exchange in the diagram of Fig.1b.
They are of the
order $(\alpha_s/\pi)$ times the r.h.s. of eq.(21)
and we will disregard
them, since they are of the same order as
perturbative corrections to
eq.(21), which are neglected.

The next term in the OPE for $\Pi_\mu(p,q)$ is
proportional to the mixed
quark-gluon condensate of dimension 5:

\begin{equation}
-g < 0 \mid \bar{q} \sigma_{\mu \nu}
G^n_{\mu \nu} \frac{\lambda^n}{2} q
\mid 0 > \equiv m^2_0 < 0 \mid \bar{q}q \mid 0 >,
\end{equation}

Consider again first the case, when the
scattering proceeds on $u$-quarks.
The calculations will be performed in the fixed point gauge

\begin{equation}
x_{\mu} A^n_{\mu}(x) = 0
\end{equation}
and we choose as a fixed point the left-side
lower vertex point in Figs.1,2.
The first source of appearance of quark-gluon
condensate is the expansion of
quark condensate in the diagram of Fig.2a in
powers of $x$. We write now

$$< 0 \mid T\{u^a_{\alpha}(x),~~ \bar{u}^b_{\beta}(0)\}
\mid 0 > = -(1/12)
\delta^{ab} \delta_{\alpha \beta}
< 0 \mid \bar{u}(0) u(0) \mid 0 > +$$
\begin{equation}
+ \frac{1}{3 \cdot 2^7} \delta^{ab}\delta_{\alpha \beta} g
< 0 \mid \bar{u}(0) \sigma_{\mu\nu} G^n_{\mu \nu}
\lambda^n u(0) \mid 0 > x^2
\end{equation}

The first term in the r.h.s. of eq.(24) gives the
already accounted quark
condensate contribution. After substituting
the second term in (24) into
the diagram of Fig.2a we obtain for the coefficient function at the
structure $p_{\mu} \hat{q} \gamma_5$

\begin{equation}
Im \Pi_{\mu} = \frac{1}{4 \pi \nu} \frac{m^2_0}{p^2}
< 0 \mid \bar{u}u \mid
0 > \frac {x}{2} p_{\mu} \hat{q} \gamma_5
\end{equation}
The structure $q_{\mu} \hat{q} \gamma_5$ is
absent in twist 2 terms. In
other way the quark-gluon condensate emerges
from the diagrams of Fig.3. To
calculate these diagrams we used the quark propagator in the constant
gluonic field \cite{4}:

$$S(x,z) = \frac{i}{(2 \pi)^4} \int\limits d^4k e^{-ik(x-z)}
\{ \frac{\hat{k}}{k^2} -
\frac{1}{4} g \lambda^n G^n_{\alpha \beta}
\varepsilon_{\alpha \beta \sigma
\rho} \gamma_5 \gamma_{\rho} \frac{k_{\sigma}}{k^4} +$$
\begin{equation}
+ \frac{1}{4} g \lambda^n G^n_{\alpha \beta}
z_{\beta}(\gamma_{\alpha} k^2 -
2 k_{\alpha} \hat{k}) \frac{1}{k^4} \}
\end{equation}
The results of the calculations of the
Fig.3 diagrams are the following. The
diagrams with gluon emission from both horizontal lines of $u$ and
$d$-quarks (Figs.3a,b) give a vanishing
contribution to the coefficients
at the structures we are interested in. The sum
of the diagrams Figs.3c,d with
gluon emission from vertical $u$-quark lines,
when the second term in the
r.h.s. of (26) is accounted, gives

\begin{equation}
Im \Pi_{\mu} = -
\frac{1}{4 \pi \nu} \frac{m^2_0}{p^2} < 0 \mid \bar{u}u
\mid 0 > \frac{1}{6} (\frac{1-x}{x}) p_{\mu} \hat{q} \gamma_5
\end{equation}
The most complicated is the calculation
of the third term in the r.h.s. of
the (26) contribution. In the chosen fixed
point this term contributes only
to the Fig.3d diagram and the calculation gives:

\begin{equation}
Im \Pi_{\mu} = - \frac{1}{4 \pi \nu}
\frac{m^2_0}{p^2} < 0 \mid \bar{u}u
\mid 0> \frac{1}{12} p_{\mu} \hat{q} \gamma_5
\end{equation}

For all Fig.3 diagrams the coefficient function at the structure
$\hat{q}q_{\mu} \gamma_5$ vanishes in twist 2 terms.
The total contribution
of dimension 5 mixed quark-gluon condensate
to the four-point correlator,
when the scattering proceeds on $u$-quarks,
is obtained by summing
(25),(27),(28)

\begin{equation}
Im \Pi^u_{\mu} = - \frac{1}{4 \pi \nu}
\frac{m^2_0}{p^2} < 0 \mid \bar{u}u
\mid 0 > \frac{1}{6}(\frac{1}{x}
- \frac{1}{2} - 3x)p_{\mu} \hat{q} \gamma_5
\end{equation}
After borelization we have

\begin{equation}
{\it B}_{M^2} Im \Pi^u_{\mu}(p,q)
= \frac{1}{4 \pi \nu} m^2_0 < 0 \mid
\bar{u}u \mid 0 > \frac{1}{6}( \frac {1}{x}
- \frac{1}{2} - 3x) p_{\mu}
\hat{q} \gamma_5
\end{equation}
For the case of scattering on
$d$-quark the contribution of quark-gluon
condensate vanishes for the same reason as the quark condensate
contribution. Since in the scattering on
$d$-quark the first two terms in
OPE vanish and the charge square of
$d$-quark is $4$ times smaller than
$u$-quark, we can safely disregard the
$d$-quark contribution to the proton
structure function $h_1(x)$.

\vspace{5mm}
\section{The sum rule. Determination of $h_1(x)$.}
\vspace{3mm}

In accordance with the results of Sec.3 consider
the coefficient function
$H(p^2,x)$ at the structure $p_{\mu} \hat{q} \gamma_5$
in the four-point
correlator (12) as a function of $p^2$.
Using the double dispersion relation
represent $H(p^2,x)$ in terms of the
contribution of physical states - the
proton and excited states

\begin{equation}
H(p^2,x) = - \frac{\pi m}{\nu}
\frac{\lambda^2_N}{(p^2-m^2)^2} h_1(x) +
\frac{\pi m}{\nu} \frac{A(x)}{p^2-m^2} + \int\limits_{W^2}^{\infty}
\frac{\rho(p^2_1,x)}{(p^2_1-p^2)^2} dp^2_1
\end{equation}
The factor $-\pi$ in the first term in (31) comes from
(4) and the factor
$m/\nu$ emerges from comparison of (18) and (19).
The last term in (31)
corresponds to the contribution of excited states
approximated by continuum.
In our case this contribution is determined
by the diagram of Fig.1a and is
given by eq.(20), $W^2$ is the continuum threshold.
The second term in (31)
is a nondiagonal term where the current
$\bar{\eta}$ produces a proton from
the vacuum, which, after interaction with
$j_{\mu5}$ and $j$ goes into some
excited state $N^*$ absorbed finally by the current
$\eta$ (for details see
\cite{4}). Apply the Borel transformation
to (31) and equate it to the
result of QCD calculations (21),(30).
We obtain the sum rule for $u$-quark
contribution to the proton structure function $h^p_1(x)$

\begin{equation}
\frac{h^u_1(x)}{M^2} + B(x) = 2a
\frac{e^{m^2/M^2}}{\tilde{\lambda}^2_N m}
[2M^2x E_1(\frac{W^2}{M^2})L^{-4/9}
\end{equation}
$$
+ \frac{m^2_0}{6}(\frac{1}{x} - \frac{1}{2} -3x)L^{-8/9}],
$$
where $B = A/\lambda^2_N$,

$$a = -(2 \pi)^2 < 0 \mid \bar{q}q \mid 0 > = 0.55 GeV^2$$
\begin{equation}
\tilde{\lambda}^2_N = 32 \pi^4 \lambda^2_N = 2.1 GeV^6,
{}~~~W^2 = 2.3 GeV^2
\end{equation}
$$E_1(z) = 1 - e^{-z}(1+z)$$
and powers the of factor $L =ln(M/\Lambda)/ln(\mu/\Lambda)$
take into account
anomalous dimensions of currents in the
correlator and operators in the OPE.
( see \cite{5,6} for details).
Note that the anomalous dimension of the scalar current
cancels with the corresponding anomalous dimension of the quark
condensate operator in the first term of (33),
whereas the anomalous dimension
of the quark-gluon condensate operator is small and we neglect it.
We assume $\Lambda=150 MeV$ and $\mu=0.5 GeV$.
The numerical values in (33) correspond to the best
description of nucleon
mass and magnetic moments in QCD sum rule approach
\cite{6,7}. In order to
get rid of unknown contribution of $B(x)$ let us
differentiate (32) over
$1/M^2$. We get

$$h^u_1(x) = 2\frac{a}{m \tilde{\lambda}^2_N}
e^{m^2/M^2} \{ 2 M^2 x
[(m^2-M^2) E_1(\frac{W^2}{M^2})
+ \frac{W^4}{M^2} e^{-W^2/M^2}]L^{-4/9}+$$
\begin{equation}
+ \frac{1}{6} m^2_0 m^2(\frac{1}{x} - \frac{1}{2} - 3x)L^{-8/9} \}
\end{equation}

\vspace{5mm}
As is seen from (32),(34), these
expressions are not correct at $x
\rightarrow 0$, the OPE breaks down at small
$x$, as is expected from
general grounds \cite{4}. In order to find the
values of $x$, where we can
believe our results, compare the magnitude of the
second term of OPE with
the first one. We find that at $x \geq 0.4$ the
second term of OPE comprises
less than 20-25\% of the total in eq.(34) and less
than 15-20\% in  eq.(32).
Therefore we may have a confidence in our results
at $x > 0.3$. At $x$ close
to 1 our results are not reliable because it is a
resonance region (see ref.
\cite{4} for more detailed explanation). In Fig.4
we plot the Borel
parameter $M^2$ dependence of $h^u_1(x)$ determined
according to eq.(34). As
is seen from the Figure, the dependence is rather weak
and this necessary
condition for the QCD sum rule validity is fulfilled.
Our basic result - the
$u$-quark contribution to proton structure function
$h^u_1(x)$ determined
according to (34), is plotted as a solid curve in
Fig.5 (at $M^2 = 1 GeV^2$).

It must be mentioned that differentiation over
$1/M^2$, which was used in
order to get (34) from (32), spoils the accuracy
of the sum rule, since the
role of continuum and of unaccounted higher order
terms of OPE increases.
Particullarly, the continuum contribution, which
comprises 30\% in eq.32 is
about 70\% or even more, especially at large
$x$ in eq.(34). (these values
refer to $M^2 = 1 GeV^2$). For this reason
and also because we accounted
only two terms of OPE, the accuracy of our results is not high.
We estimate
it by $ \sim 30\%$ at $x \approx 0.5$ and by 50\% at the
border of interval
where our approach is working, $x = 0.3$ and $x = 0.7$ .
At $x > 0.7$ the
curve of $h^u_1(x)$ given in Fig.4 is not correct,
since it violates the
inequality $\mid h^u_1(x) \mid < u(x)$ proved
in ref. \cite{2}.

As was explained above, the contribution of $d$-quarks is
small and the proton structure function

\begin{equation}
h^p_1(x) \approx (4/9)h^u_1(x).
\end{equation}

In Fig.5 we also plotted $h^u_1(x)$ found by Jaffe and Ji in
the bag model
\cite{2} (dashed curve). The bag model curve is
2-3 times higher compared
with the results of our calculations and strongly
depends on $x$ at
intermediate $x$, while our $h^u_1(x)$ is flat
(due to errors we cannot
insist on negative curvature of $h^u_1(x)$,
probably it is an artifact of
our approximations). It should be mentioned that
the bag model curve
strongly violates the inequality $h^u_1(x) < u(x)$,
if one uses the
experimental values for $u(x)$.

In order to establish an input for comparison
with data of future
experimental studies as well as for possible renormalization
group analysis we suggest the following ansatz for $h^p_1(x)$ in the
whole region of Bjorken variable, based on our calculation.
We recall first that the behaviour of structure functions at
$x \rightarrow 0$
is governed by the Regge trajectory in the t-channel of the
corresponding forward scattering amplitude ( see for details,
\cite{IoffKhLi}) so that, in particular,
\begin{equation}
h^p_1(x) \sim x^{-\alpha_{a_1}(0)}
\end{equation}
where $\alpha_{a_1}(0)$ is the intercept of the axial
$a_1$-meson trajectory.
Assuming that this trajectory is linear and has the same slope as
$\rho, a_2 $ trajectories, namely $\alpha'\simeq 1GeV^{-2}$,
and using the
mass of the  $a_1$ meson we easily get
$\alpha_{a_1}(0) \simeq - 0.3 $.
We then extrapolate $h^p_1(x)$ from the point $x=0.3$,
the minimal value
at which our calculation is reliable, to the region $0<x<0.3$.
using  our result for $h^u_1(x=0.3)$ and (35),(36).
Note that Regge behaviour was also invoked in
\cite{2} for analysis of the moments of $h_1(x)$.

In the large $x$ region where the structure function should decrease
according to the general quark counting rules,
we simply use the inequality
$u(x)>h_1^u(x)$ established in \cite{2} and replace
$h_1^u(x)$ by the corresponding values of the $u(x)$ distribution
( taking it, e.g. from \cite{GRV} )beginning from the point $x=0.55$
where $h_1^u(x)$ starts to violate this
inequality. ( Within the uncertainties of
$h_1^u(x)$ and $u(x)$ this point
may be as large as $x \simeq 0.7$, the upper
limit of the interval of
reliability of our calculations). The resulting curve
is shown in Fig.6.

Finally, let us compare our predictions for $h_1(x)$ with the
nucleon spin structure function $g_1(x)$.
Using recent data of SLAC E142 experiment \cite{SLAC} on the neutron
spin
structure function  $g_1^n(x)$  measured at
$Q^2 >1GeV^2$ we conclude that at
$x>0.3$ this structure function values are consistent with zero.
In terms of separate $u$ and $d$ quark spin
distributions in the proton,
the following relation should be valid  starting from $x=0.3$ :
$g_1^u(x) \simeq -4g_1^d(x)$ clearly indicating that the
the proton spin distribution  $g_1^{p}(x)$, as well as $h_1^p(x)$, is
dominated at $x > 0.3$ by the u-quark contribution:
\be
g_1^{p}(x) \simeq (4/9)g_1^{u}(x)~.
\label{49gu}
\ee
Taking experimental results on $g_1^{p}(x)$ from EMC
measurement \cite{EMC}
we get, e.g. $g_1^{u}(x) \simeq 0.5~ (0.2)$ at $x=0.3~ (x=0.5)$.
The recent SMC measurement \cite{SMC} gives the values of
$g_1^u(x)$ at $x=0.3$ and $0.5$ about $20 \%$ lower, but agreeing with
EMC within errors.
Comparison with our calculation clearly indicates the validity of
the inequality for $u$ quark distributions in the proton:
\be
h_1^u(x) > g_1^{u}(x)
\label{ineq}
\ee
at $x \geq 0.3$.
The same conclusion can be obtained, if we compare $h_1(x)$ with
the structure function $g_1(x)$ calculated in \cite{g1} in the framework
of the method, similar to used here. Unfortunately, the $g_1(x)$
calculation is legitimate
only in the narrow domain of $x$, $0.5 < x< 0.7$. In this domain the
results of \cite{g1} are in good agreement with recent data of SMC
\cite{SMC}. In the interval $0.4<x<0.7$ the result of this experiment
for the mean value of the proton spin structure function is
$\overline{g_1^p(x)} =0.08\pm0.02\pm 0.01$
(see Table 1 in \cite{SMC}). The prediction of ref. \cite{g1}
for the same interval of $x$ was
$\overline{g_1^p(x)}=0.06$ with an estimated error about 30$\%$
(some extrapolation of the $g_1^p(x)$ curve from $x=0.5$ to $x=0.4$ is done).
As was shown in \cite{g1}, the u-quark contribution to the spin structure
function dominates in this region of $x$, in accordance with ($\ref{49gu}$).
Using the results of \cite{g1} we can compare theoretical predictions of
$h_1^u(x)$ and $g_1^u(x)$ at the point $x=0.5$ (but not at $x=0.3$).
We find from \cite{g1} $g_1^u(0.5)\simeq 0.14$  which agrees
within theoretical and experimental errors with the value obtained above
from EMC experiment and with inequality (\ref{ineq}).
The results of bag model calculations \cite{2} also obey
(\ref{ineq}).

\vspace{7mm}
\section{Conclusion}
\vspace{5mm}
In this paper a new method of calculation of the proton chirality
violating structure function $h^p_1(x)$
at intermediate $x$ has been developed based on the QCD sum rule approach.
The obtained estimates for this structure function are valid at ~ $0.3 <
x<0.7$.
Since we disregard the perturbative QCD correction, our results are
valid at intermediate $Q^2 \approx 5-10 GeV^2$, where deviations from
scaling are inessential. In our approach $h^p_1(x)$ is proportional
to the quark condensate. We have calculated only two terms in the OPE.
Nevertheless, once the method is established, the accuracy may be
improved in future taking into account higher order terms of OPE. We estimate
the current uncertainty of our calculation at the level of 30-50\%.
Therefore, having in mind that the similar calculation of the proton spin
structure function $g_1(x)$ has reasonable agreement with experiment
we consider our estimate of $h^p_1(x)$ as a reliable and
useful guideline for experiments aimed at measuring
this structure function.

We found that $h^p_1(x)$
is rather flat in the interval $0.3 < x < 0.7$ and not
too much smaller than
an upper limit $\mid h^u_1(x) \mid < u(x)$ derived in \cite{2}.
This
circumstance  gives a good chance for its experimental study.
Another
interesting feature of $h^p_1(x)$ is that the main
contribution comes from
$u$-quarks,
the $d$-quark contribution
being suppressed, besides of the $d$-quark charge,
by a factor of order
$\alpha_s/\pi \sim 0.1$ or is given by the
contribution of higher dimension
operators in OPE, which, probably, are also small.
For this reason we expect
that for the neutron at intermediate $x$
$$
h^n_1(x) \approx (1/4) h^p_1(x)
$$

The comparison of our $h^u_1(x)$ with the same function
calculated by Jaffe
and Ji in the bag model \cite{2} shows that our
$h^u_1(x)$ is by a factor
of 2-3 smaller in the interval $0.3 < x < 0.5$ and
matches with $h^u_1(x)$
at $x = 0.6$. The condition
$\mid h_1(x) \mid > \mid g_1(x) \mid$ found in
the bag model \cite{2} is also fulfilled in our calculation.

{\bf Acknowledgements}.
One of the authors (B.I.) is thankful to R. Jaffe for the suggestion of
eq.(2) and for the hospitality at MIT. A.K. acknowledges
support from the
A.von Humboldt Foundation.

\end{document}